# Electroabsorption spectroscopy of single walled nanotubes

J. W. Kennedy and Z. V. Vardeny*

*Physics department, University of Utah Salt Lake City Utah 84112*

S. Collins and R.H. Baughman

*Nano Tech Institute, University of Texas at Dallas, Richardson, Texas 75083*

and

H. Zhao and S. Mazumdar

*Physics Department, University of Arizona, Tucson, Arizona 85721*

Abstract

We have measured the electric field modulated absorption of a sample of single-walled nanotubes (SWNT) suspended in a solid polyvinyl alcohol matrix. The electroabsorption (EA) spectrum roughly follows the first derivative of the absorption with respect to photon energy, scales quadratically with the electric field strength, and shows a pronounced anisotropy of light polarization with respect to the applied electric field direction. These findings indicate a quadratic Stark effect caused by a change in the polarizability of the excited states, which is common to quasi-one dimensional (1D) excitons in organic semiconductors. The EA spectrum is well described by calculations involving electron-electron interaction in the model Hamiltonian of both zigzag and chiral nanotubes. We have calculated the EA spectra for both zigzag and chiral nanotubes within a model Hamiltonian that includes electron-electron interactions. The calculations reproduce the observed quadratic Stark shift of the lowest optical exciton, as well as the more complicated behavior of the EA spectrum in the energy region that corresponds to the next higher exciton. Our findings show that the low-lying absorption bands in semiconducting SWNT are excitonic in origin, in agreement with transient optical measurements that identify the primary photoexcitations in SWNT as quasi-1D excitons with a substantial binding energy.

*Author to whom correspondence should be addressed; e-mail: val@physics.utah.edu

1. Introduction

The optical properties of carbon nanotubes (NTs), and in particular single-walled carbon nanotubes (SWNTs) have received recently extensive attention. Photoluminescence and resonant Raman scattering spectroscopies have emerged as powerful methods of characterizing the diameter and chirality of individual SWNT in solutions and thin films [1-5]. Studies involving ultrafast transient spectroscopies [6, 7], as well as non-linear optical spectroscopies [8] applied to NT films and solutions have begun to elucidate the nature of the primary photoexcitations in bundled and isolated SWNTs. Also NT field-effect transistors were used to reveal the intricate mechanism of charge photogeneration in individual SWNTs [9]. In spite of these spectroscopies applied to NTs an outstanding question still remains, namely whether the observed optical transitions are excitonic or interband transition in nature. In the simplest interpretation semiconducting NT (*S*-NT) is a direct band gap material, and thus the lower optical transition energy should give the NT band-gap, which is one of the most important quantities for semiconductors. However, NTs are quasi-one dimensional (1D) confined materials with a typical diameter of ~1 nm, which leads to relatively strong electron-hole (e-h) Coulomb interaction, and consequently to strongly bound excitons [10].

Recent state of the art electronic structure calculations [11-14] predict the existence of strongly bound excitons in small diameter NTs, with a binding energy of the order of 0.5 eV that decreases with the NT diameter [14]. Based on these experimental and theoretical studies a consensus emerges that although the earliest description of the NT electronic structure was treated within one-electron theory, electron-electron and e-h interactions play a significant role in these systems, and consequently the primary photoexcitations in *S*-NT are relatively strongly bound excitons [6, 7] rather than free electrons and holes. Within the interaction picture, corresponding to each pair of quantized valence and conduction bands, there occurs a quasi-1D optical accessible exciton and higher lying energy states that include a continuum band. The overall electronic structure of a *S*-NT consists therefore of several such energy manifolds (that has been labeled as n = 1, 2, etc.), with each manifold containing both excitons and continuum bands [10, 14].

To elucidate the nature of the optical absorption bands in NTs we have chosen to apply electroabsorption (EA) spectroscopy. EA has provided a sensitive tool for studying the band structure of inorganic semiconductors [15], as well as their organic counterparts [16-19]. Transitions at singularities of the joint density of states respond particularly sensitively to an external field, and are therefore lifted from the broad background of the absorption continuum. The EA sensitivity decreases, however in more confined electronic materials, where electric fields of the order of 100 kV/cm are too small of a perturbation to cause sizable changes in the optical spectra. As states become more extended by inter-molecular coupling they respond more sensitively to an intermediately strong electric field, F since the potential variation across such states cannot be ignored compared to the separation of energy levels. EA thus may selectively probe extended states and thus is particularly effective for organic semiconductors, which traditionally are dominated by excitonic absorption. One of the most notable examples of the application of EA spectroscopy to organic semiconductors is polydiacetylene, in which

EA spectroscopy was able to separate absorption bands of quasi-1D excitons from that of the continuum band [20]. The confined excitons were shown to exhibit a quadratic Stark effect, where the EA signal scales with $F^2$ and the EA spectrum is proportional to the derivative of the absorption respect to the photon energy. In contrast, the EA of the continuum band scales with $F^{1/3}$ and shows Frank-Keldish (FK) type oscillation in energy. The separation of the EA contribution of excitons and continuum band was then used to obtain the exciton binding energy in polydiacetylene, which was found to be ~ 0.5 eV [20]. Since NTs are quasi-1D organic semiconductors it is thus natural to apply the EA spectroscopy to these materials.

Here for the first time we found that the EA spectrum of well-separated NTs embedded in a transparent polymer matrix show a quadratic Stark effect, having a spectrum that is similar to the first derivative of the absorption spectrum respect to energy, and a quadratic dependence on F; no FK type oscillations were observed. In addition we also found a pronounced anisotropy in EA for light polarization parallel and perpendicular to the direction of the applied electric field. These findings show that carbon NTs optics is dominated by strongly bound excitons, in contrast to the earlier assumptions. The polarization anisotropy also shows that the excitonic absorption bands are relatively confined along the individual NTs. The EA spectrum and its $F^2$ dependence is well-described by a theory involving e-e interaction in the model Hamiltonian. In particular we found in both experiment and theory that the EA spectrum of the second band in the absorption spectrum shows a phase-shift like feature as a result of strong coupling to both lower and upper lying excited states.

2. Experimental

In order to measure the optical response of SWNTs to an applied electric field it was necessary to fabricate a highly resistive, transparent sample. To achieve this we mixed 0.005% HiPCO-produced SWNTs with 0.610% SDS surfactant and 0.865% poly-vinyl alcohol (PVA) in de-ionized water. Sonication for an hour before sample preparation resulted in relatively well-separated NTs as evident by the sharp features in the absorption spectrum (Fig. 1). We then deposited a film of the solution onto our electroabsorption substrate by drop-casting at 80º C. The EA sample consisted of well-separated SWNTs embedded in an electrically insulating matrix of PVA having an optical density of ~ 1 in the visible/near ir spectral range (Fig. 1). Neither PVA nor SDS has absorption bands in the spectral range over which we measured the EA spectrum. Resonant Raman scattering of the radial breathing mode was used to determine that the NTs in our sample have a diameter distribution around a mean diameter of ~ 0.8 eV, and contain about 1/3 metallic and 2/3 S-NTs.

The EA substrate consisted of two interdigitated sets of a few hundred gold electrodes 30 μm wide patterned on a sapphire disk 1mm thick. An electric field was generated in the sample by applying a potential V to the electrodes. An applied potential V = 600 Volts (typical for our experiments) resulted in field strength $F = 2\times10^5$ V/cm. We varied V on the electrodes using a sinusoidal signal generator at f = 1 kHz, and a simple transformer to achieve high voltages. The modulation ΔT of the transmission T is expected to be the same for positive and negative V's, since the NTs are not preferentially aligned with respect to the electrodes. Thus, we measured ΔT using a diode detector and a lock-in amplifier set to twice the frequency (2f) of the applied field [19]. We verified that no EA signal was observed at f or at 3f. ΔT and T spectra were measured separately using a homemade spectrometer that consisted of a ¼ meter monochromator equipped with several gratings and solid-state detectors such as InSb, Ge, Si and Si-UV enhanced diodes to span the EA in the broadest spectral range of 0.2 to 3.0 eV. The EA spectrum was obtained from the ratio ΔT/T, which was measured at various applied voltages and polarizations of the probe light respect to the direction of the applied field.

3. Experimental Results

The optical absorption spectrum $\alpha(E)$ of the SWNT/PVA sample from 0.5 eV to 3 eV along with the numerically calculated spectrum of the first derivative of the absorption respect to energy E, $d\alpha/dE$ is shown in Fig. 1. We observed distinct absorption peaks in each of the regions traditionally labeled as the 1-1 (around 0.8 eV) and 2-2 (around 1.7 eV) Van Hove singularities transitions for NTs of ~ 0.8 nm in diameter [21]. The absorption spectrum is thus in agreement with having a distribution of different NTs in the sample, and thus inhomogeneity plays an important role in the interpretation of the spectroscopic data.

The EA spectrum obtained in the same spectral range by applying an electric field $F = 10^5$ V/cm is shown in Fig. 2; $d\alpha/dE$ spectrum is also shown for comparison. With the exception of a few differences to be discussed hereafter, the two spectra are quite similar. The spacing between pairs of adjacent peaks is the same for the two spectra within 0.02eV. This is a surprising result when considering the broad distribution of NTs in the sample, and their division into semiconducting and metallic NTs that should have masked the simple relation between the two spectra, if the contribution to the EA of different NTs would vary strongly among them. We therefore conclude that the contribution of individual NT to EA is not a strong function of its chirality, diameter or whether it is metallic or semiconducting in nature. Whereas the EA and $d\alpha/dE$ spectra match very well to each other with a slight blue shift of the EA spectrum for energies E < 1.4 eV, we observe that in the energy region E > 1.4 eV it seems that the EA signal changes sign respect to that of $d\alpha/dE$. Such a phase shift between EA and $d\alpha/dE$ spectra is exactly what is found in our calculated EA spectra in the region of the second exciton in chiral NTs (see Figs. 5(b) and 6(b) and the discussion below). In addition, the EA spectrum also contains a positive absorption contribution below 1.4 eV that is absent in the $d\alpha/dE$ spectrum. We note that the EA signal becomes noisy above about E = 2.2 eV because of reduced transmission through the sample caused by higher absorption in the film. In spite of this it is apparent that the EA signal at E > 2.2 eV becomes smaller with increasing photon energy.

The EA spectrum between 0.8 eV and 1.3 eV is shown in Fig. 3 for 5 different field strengths. The inset shows the obtained linear dependence of the EA signal on $F^2$ for four different peak positions in the spectrum. Whereas the dependence on $F^2$ is different for different regions of the spectrum, each region shows a quadratic dependence on F. The $F^2$ dependence is consistent with a quadratic Stark shift. In addition we do not find in the entire spectral range any nonlinear spectral oscillation in energy that would be indicative of FK EA oscillation.

To investigate the extended wave-function associated with the transitions in the NT absorption spectrum we also measured the EA polarization dependence. The EA spectra for three different polarizations of probe light respects to the direction of the applied field are shown in Fig. 4. The EA shows a strong anisotropy with respect to the linear polarization of the probe light. The spectra are more pronounced when the light is polarized parallel to the applied field. The ratio of the signal at 0º angle between the light polarization and field direction, to the signals at 45º and 90º angles is ~ 1.3 and ~ 3.2, respectively. This anisotropy shows that the EA signal is related to a confined excitation along the NT direction; otherwise this anisotropy would be washed out for extended excitations. The EA anisotropy agrees well with other polarized optical measurements such as polarization anisotropy in PL and picosecond transient optical spectra recently completed [2, 5].

4. Theory

The detailed theory of EA in π-conjugated polymers was formulated in ref. [18]. In the presence of an external static electronic field **F** the overall Hamiltonian of the system is given by

$$H = H_0 + e\mathbf{F}\cdot\mathbf{r} = H_0 + \boldsymbol{\mu}\cdot\mathbf{F} , \qquad (1)$$

where $H_0$ is the unperturbed Hamiltonian that describes the system in the absence of the field, and **μ** is the dipole operator. The electric field therefore introduces off-diagonal matrix elements between one- and two-photon states that are directly proportional to the field. For a quasi-one-dimensional (quasi-1D) system such as a π-conjugated polymer or a S-NT, the largest effect of the field occurs when it is along the 1D axis of the system, since the dipole matrix elements are largest along this axis. The EA for the field along the 1D axis is then calculated as $\Delta\alpha(\omega,F_z) = \alpha(\omega,F_z) - \alpha(\omega,0)$, where $\alpha(\omega,F_z)$ is the absorption of the system with field $F_z$. From Eq. (1), the effect of the field is strongest for states within a continuum band, because of the close proximity in energy between states that are coupled by the field-dependent term. Discrete excitonic one-photon allowed states exhibit a Stark shift, while previously forbidden two-photon excitonic states become weakly allowed due to admixing with previously allowed one-photon states and appear as peaks in the EA spectrum. Importantly, the Stark shift of the optical exciton in π-conjugated polymers is a redshift, as opposed to the blueshift expected within a two-level model, indicating that the field-induced coupling between the exciton and a higher energy two-photon state is larger than the coupling between the ground state and the exciton [18].

We have calculated the EA in three different S-NTs [(10,0), (6,2), (6,4)] within Eq. (1). In all cases our system Hamiltonian $H_0$ is the correlated Pariser-Parr-Pople type Hamiltonian that has been used before to demonstrate the excitonic energy of S-NTs [14]. We use the same procedure here to solve $H_0$ within the single-configuration interaction scheme. We calculate all transition dipole couplings between the excited states of $H_0$ to evaluate the field-dependent off-diagonal matrix elements of $H$. We then diagonalize $H$ and calculate $\alpha(\omega,F_z)$ for several different electric fields. In Figs. 5(a) and (b) we show the results of our calculations for the (10,0) and (6,2) *S*-NTs, respectively. In both cases the upper panel corresponds to the linear absorption, whereas in the lower panel the calculated EA is compared to the d$\alpha$/dE derivative spectrum. The calculated EA spectra show Stark shifts of the n = 1 and 2 excitons, and in addition, an oscillatory EA signature in between the signatures due to the excitons. The oscillation is ascribed to the n=1 continua from examination of the eigenstates' energies in this energy region. The absence of such oscillatory signatures in the experimental EA spectra indicates the disordered nature of the NT/PVA mixture; this is in addition to the inhomogeneity caused by the distribution of *S*-NTs with different absolute exciton energies and exciton binding energies.

Focusing on the Stark shift of the n=1 excitons, we see that in both the zigzag and chiral NTs, there is a one-to-one correspondence between the EA signature and the derivative spectrum, in agreement with the experiment. We have confirmed that the amplitude of the EA scales as $F^2$. In both (10,0) and (6,2) *S*-NT there occur two-photon exciton states above the optical exciton that are very strongly dipole-coupled to the latter. In the case of the chiral (6,2) NT, there actually occur several such two-photon exciton states, forming a narrow band. We also note that the calculated EA signal for the n=1 exciton is larger than that for the n=2 exciton for the zigzag NT and is smaller for the chiral NT.

A significant feature of the calculated EA in the energy region corresponding to that of the n=2 exciton is the relative "phase shift" between the EA signal and the derivative spectrum in the case of the (6,2) S-NT. *Such a relative phase shift is precisely what is observed experimentally in this energy region.* To clarify this further, we have shown in Figs. 6(a) and (b) the "blown up" EA signals corresponding to this energy region for the (10,0) and the (6,2) S-NT. *We have confirmed that the EA signal due to the n=2 exciton are similar for the (6,2) and (6,4) NTs.* The EA signal indicates an energy shift that is considerably smaller in magnitude than that of the n=1 exciton (with even a very weak tendency to have opposite sign for the shift, upon close examination of the numerical results), with the dominant effect of the electric field being the reduced oscillator strength of the optical transitions into this exciton. We have examined all transition dipole couplings between this exciton state and two-photon states in this energy region, and have determined the fundamental reason for this peculiar nonlinear response. Specifically, the n=2 exciton in chiral NTs are once again strongly coupled to a band of two-photon states, but unlike what happens in the n=1 energy region, this band is now rather broad, and extends both below and above the optical exciton [22]. There is thus a tendency to cancellation of the field-induced energy shift of the n=2 exciton, giving rise to the phase shift between the EA and the derivative signal. Since there ought to be many more chiral than zigzag S-NTs in the experimental material, it is not surprising that the experimental EA resembles the calculated EA for the chiral S-NTs.

5. Conclusions

In conclusion we measured for the first time the EA spectrum of well-separated SWNT embedded in a transparent PVA matrix. We found that the EA spectrum roughly resembles the d$\alpha$/dE spectrum, scales with $F^2$, and is highly anisotropic respect to the direction of the applied field; these results are consistent with a quadratic Stark effect of confined transitions. We also conclude that NTs contribute similarly to EA regardless of their chirality, and diameter within the sample narrow diameter distribution. In addition, the absence of any FK oscillation shows that the interband transition related to the continuum band is dominated by the disorder and inhomogeneity in the sample that prevents this type of EA signal to form. The experimental results and the good agreement with the applied theory that involves e-e interaction in the model Hamiltonian, strongly indicate that the absorption bands observed in SWNTs are in fact excitonic transitions,

## 6. Acknowledgements

This work was completed while one of us (ZVV) was on sabbatical leave at the Technion; support from Lady Davis foundation is gratefully acknowledged. The work at the University of Utah was supported in part by the DOE grant # FG-04-ER46109. At the University of Texas at Dallas the work was supported by DARPA grant No. MDA 972-02-C-005 and the Robert A. Welch foundation. At the University of Arizona the work was supported by NSF grant # DMR-0406604.

**References**

1. S. M. Bachilo, M. S. Strano, C. Kittrell, R. H. Hauge, R. E. Smalley, and R. B. Weisman, Science **298**, 2361 (2002).

2. A. Jorio, A. G. Souza Filho, V. W. Brar, A. K. Swan, M. S. Ünlü, B. B. Goldberg, A. Righi, J. H. Hafner, C. M. Lieber, R. Saito, G. Dresselhaus, and M. S. Dresselhaus, Phys. Rev. B **65**, 121402 (2002).

3. J. Lefebvre, J. M. Fraser, P. Finnie, and Y. Homma, Phys. Rev. B **69**, 075403 (2004).

4. H. Htoon, M. J. O'Connell, S. K. Doorn, and V. I. Klimov, Phys. Rev. Lett. **94**, 127403 (2005).

5. H. Telg, J. Maultzsch, S. Reich, F. Hennrich, and C. Thompsen, Phys. Rev. Lett. **93**, 177401 (2004).

6. O. J. Korovyanko, C.-X. Sheng, Z. V. Vardeny, A. B. Dalton, and R. H. Baughman, Phys. Rev. Lett. **92**, 017403 (2004).

7. G. N. Ostojic, S. Zaric, J. Kono, V. C. Moore, R. H. Hauge, and R. E. Smalley, Phys. Rev. Lett. **94**, 097401 (2005).

8. A. Maeda, S. Matsumoto, T. Takenobu, Y. Iwasa, M. Shiraishi, M. Ata, and H. Okamoto, Phys. Rev. Lett. **94**, 047404 (2005).

9. M. Freitag, Y. Martin, J. A. Misewich, R. Martel, and Ph. Avouris, Nano Lett. **3**, 1067 (2003); X. Qiu, M. Freitag, V. Perebeinos, and Ph. Avouris, Nano Lett. (in press).

10. T. Ando, Jour. Phys. Soc. Japan **66**, 1066 (1997).

11. C. D. Spataru, S. Ismail-Beigi, L. X. Benetict, and S. G. Louie, Phys. Rev. Lett. **92**, 077402 (2004).

12. E. Chang, G. Bussi, A. Ruini, amd E. Molinari, Phys. Rev. Lett. **92**, 196401 (2004).

13. V. Perebeinos, J. Tersoff, and Ph. Avouris, Phys. Rev. Lett. **92**, 257402 (2004).

14. H. Zhao and S. Mazumdar, Phys. Rev. Lett. **93**, 157402 (2004).

15. R. K. Willardson and A. C. Beer eds., *Semiconductors and Semimetals,* Vol. 9 (Academic Press, New York, 1972).

16. L. Sebastian, G. Weiser, and H. Bassler, Chem. Phys. **61**, 125 (1981)

17. G. Weiser, Phys. Rev. B **45**, 14076 (1992).

18. D. Guo *et al*., Phys. Rev. B **48**, 1433 (1993).

19. M. Liess, S. Jeglinski, Z. V. Vardeny, M. Ozaki, K. Yoshino, Y. Ding, and T. Barton, Phys. Rev. B **56**, 15712 (1997); and references therein.

20. L. Sebastian, and G. Weiser, Phys. Rev. Lett. **46**, 1156 (1981).

21. C. X. Sheng, Z. V. Vardeny, A. B. Dalton, and R. Baughman, Phys. Rev. B **71**, 125427 (2005).

22. H. Zhao, C.-X. Sheng, S. Mazumdar, and Z. V. Vardeny, to be published.

**Figure Captions**

Fig. 1: The linear absorption spectrum, $\alpha(E)$ and the numerically calculated first derivative $d\alpha/dE$ spectrum for a film of SWNTs embedded in a PVA matrix.

Fig. 2: The Electroabsorption spectrum (upper curve) at field strength $F = 10^5$ V/cm compared to $d\alpha/dE$ spectrum (lower curve). The two curves were displaced for ease of comparison.

Fig. 3: The EA spectrum at five different field strengths. The inset shows the dependence of EA on $F^2$ for the four peaks seen in the spectrum.

Fig. 4: The EA spectrum at three different probe light polarizations with respect to the direction of the applied electric field in the film.

Fig. 5: (a) The calculated linear absorption (upper panel) and the EA (lower panel) spectra for the (10,0) *S*-NT, within the PPP model using the parameters of ref. [14]. All energies are in units of the nearest neighbor hopping integral, t. The EA spectrum is compared to the $d\alpha/dE$ spectrum in the lower pane. (b) Same as in (a) but for the (6,2) chiral *S*-NT.

Fig. 6: Same as in Fig. 5, but in the energy region of the n = 2 exciton of the (10,0) (a) and (6,2) chiral *S*-NT (b). Note the relative shift between the EA and $d\alpha/dE$ spectra for the (6,2) chiral NT.

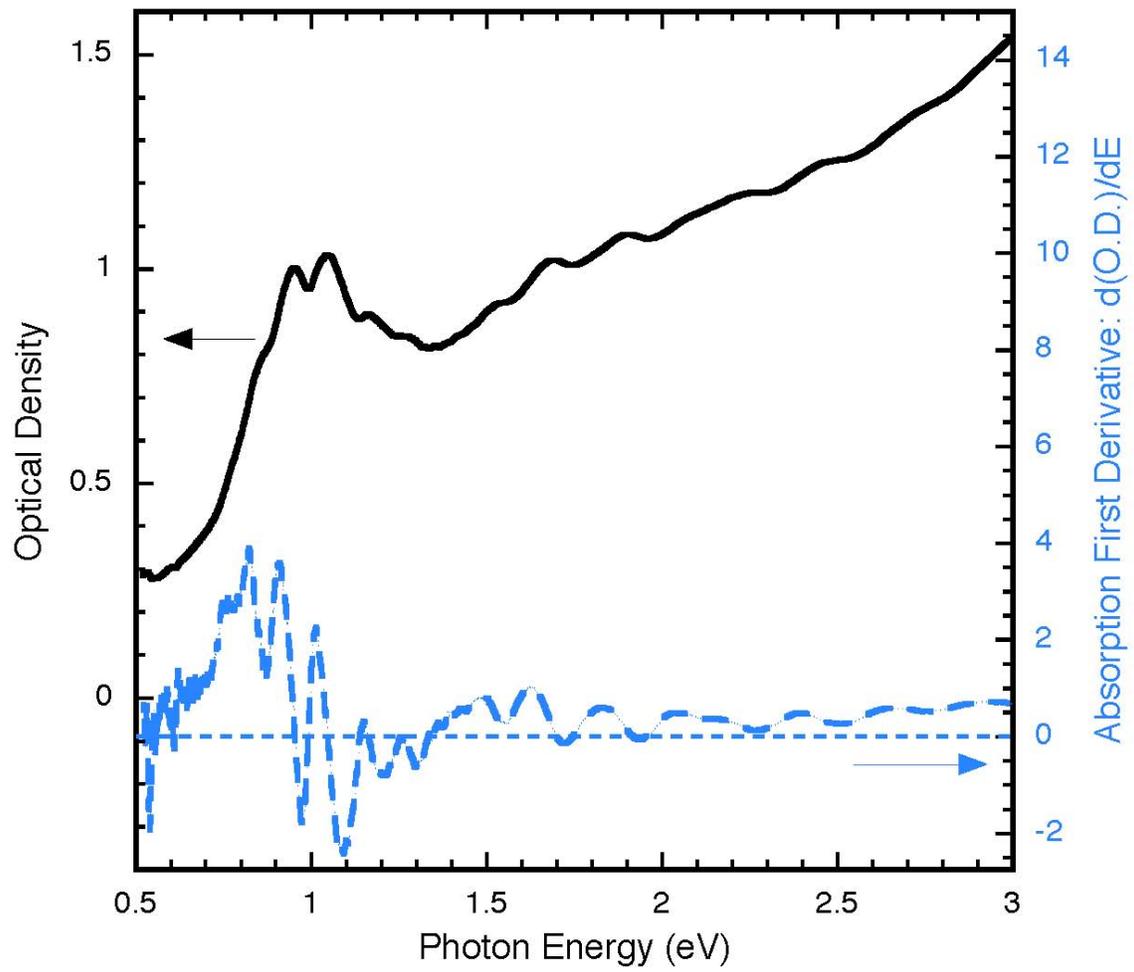

Fig. 1

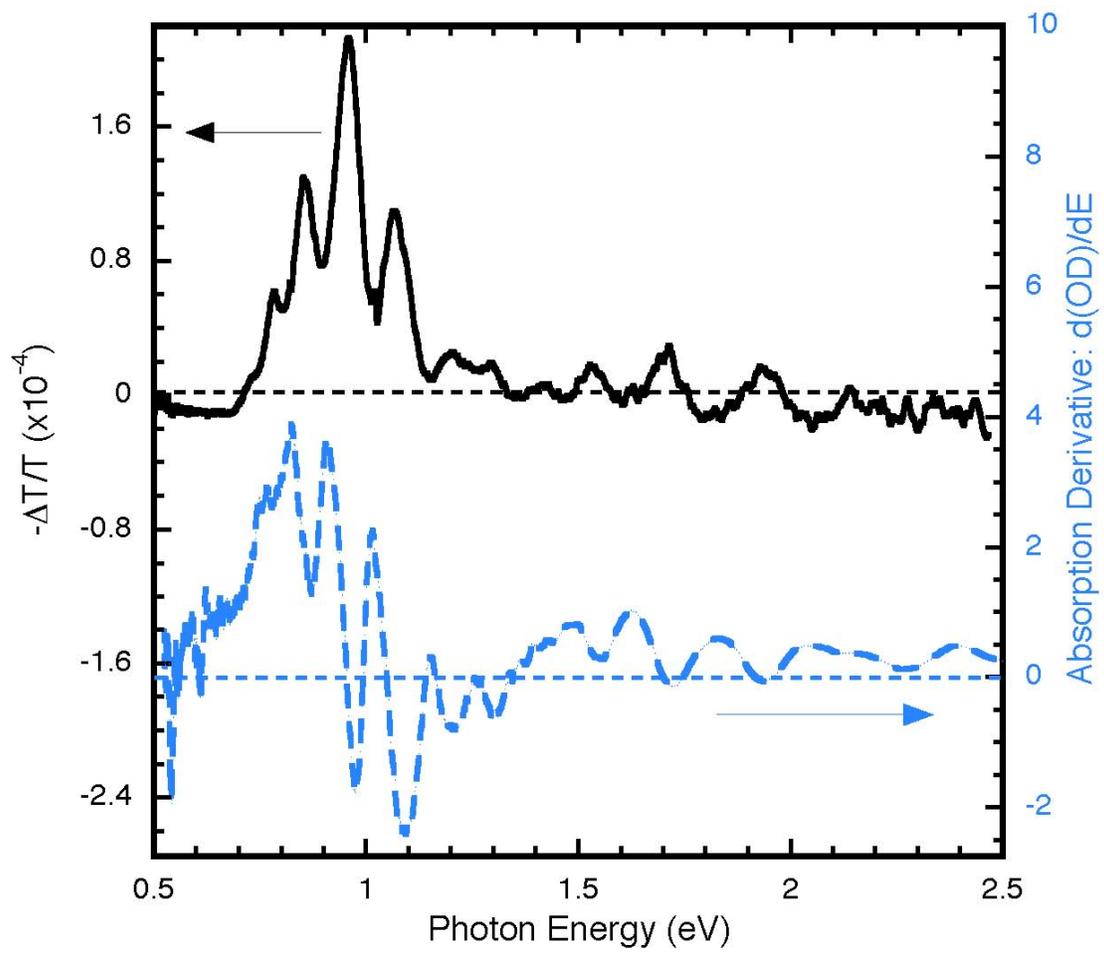

Fig. 2

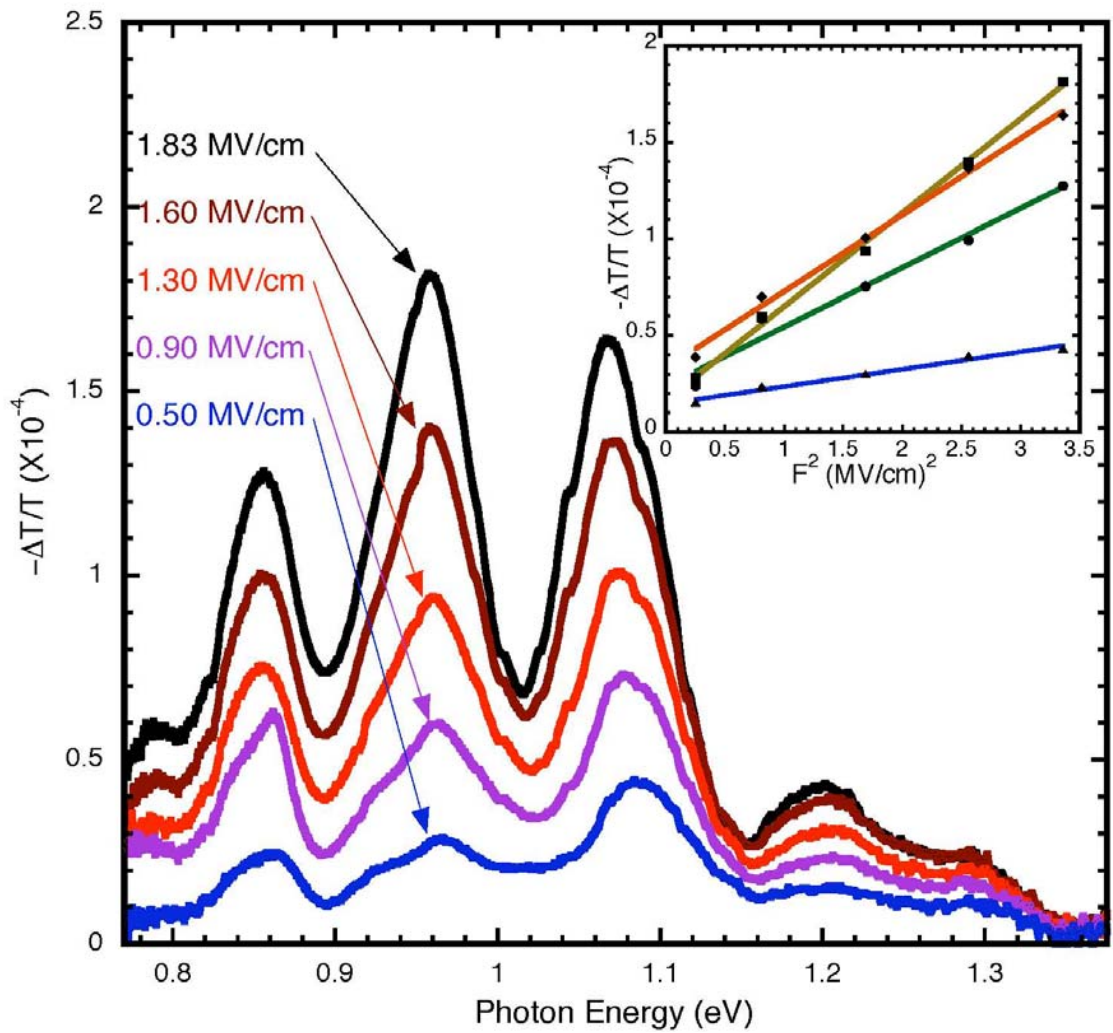

Fig. 3

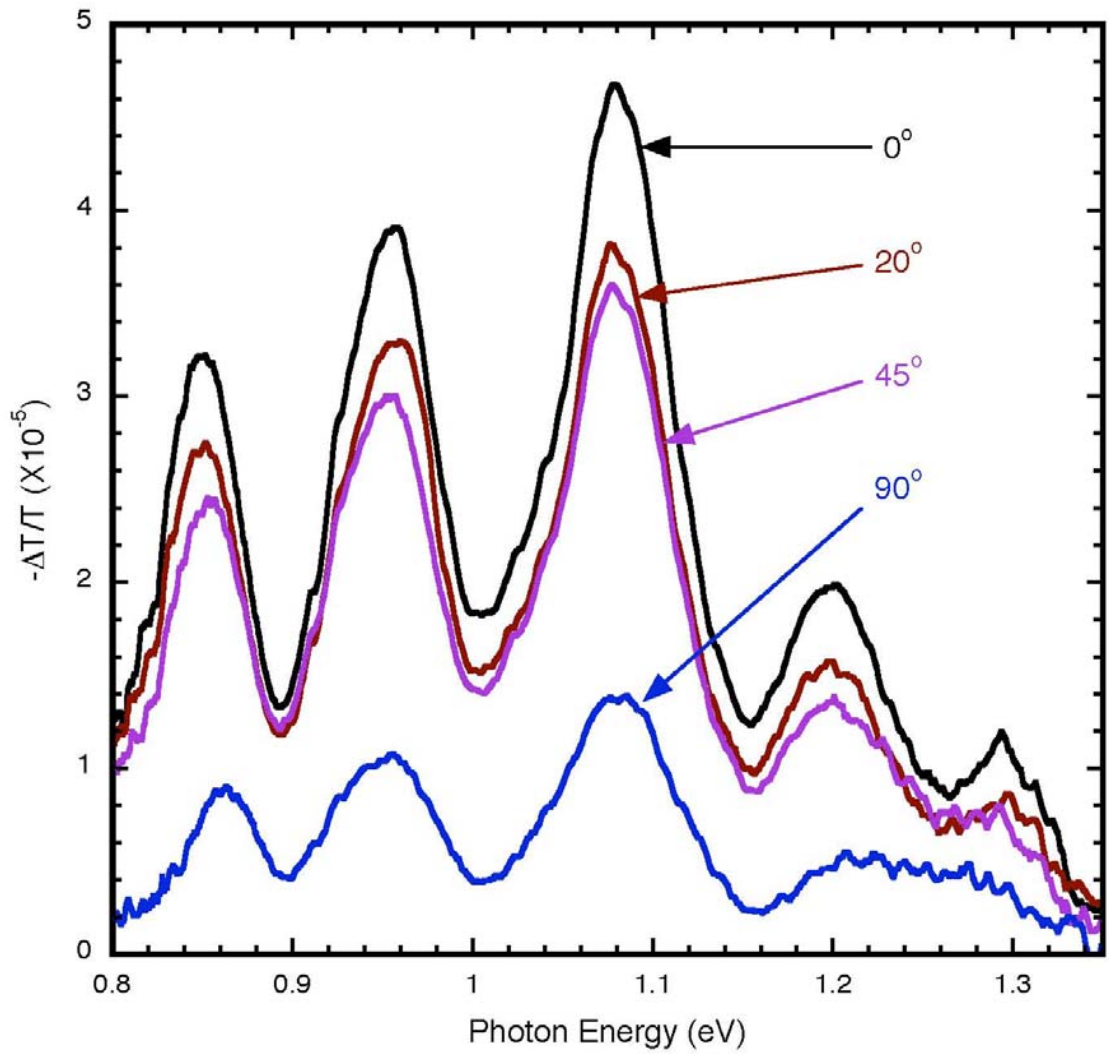

Fig. 4

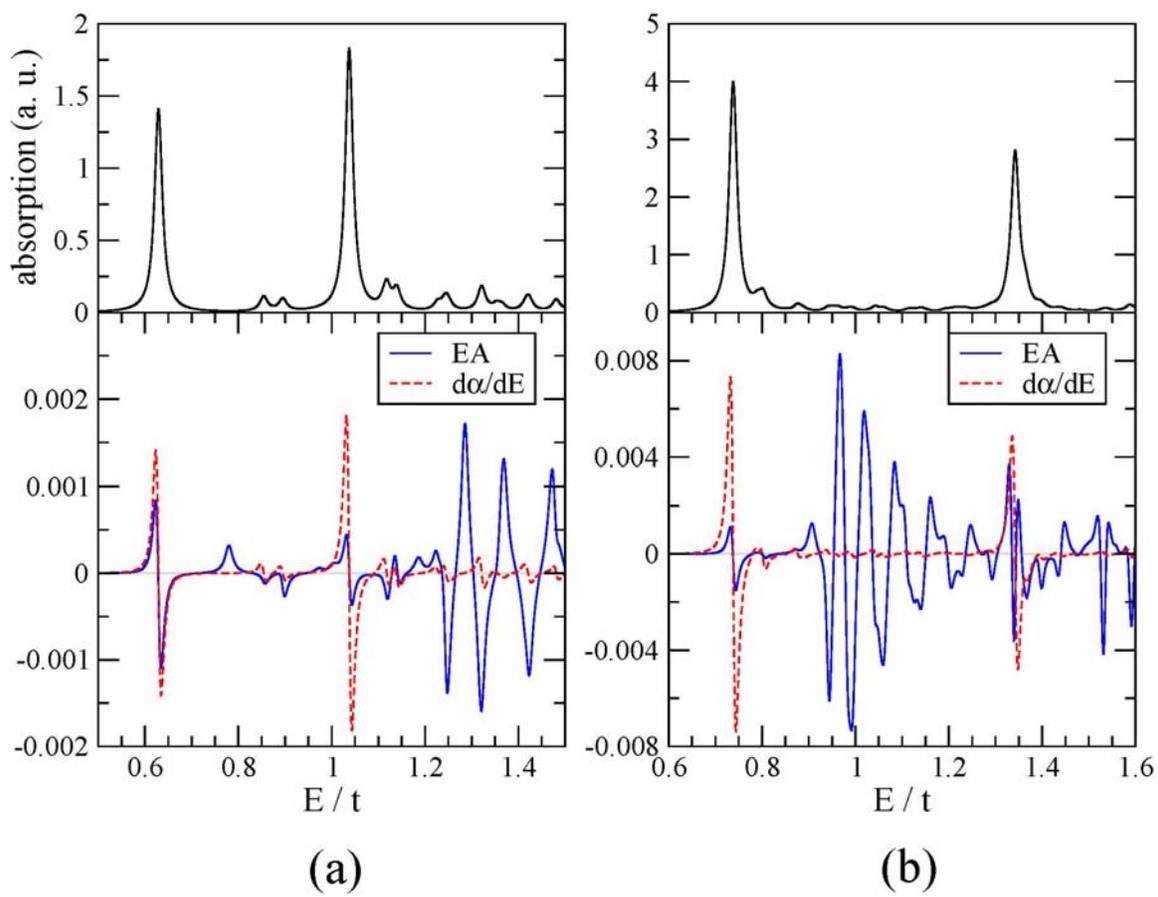

Fig. 5

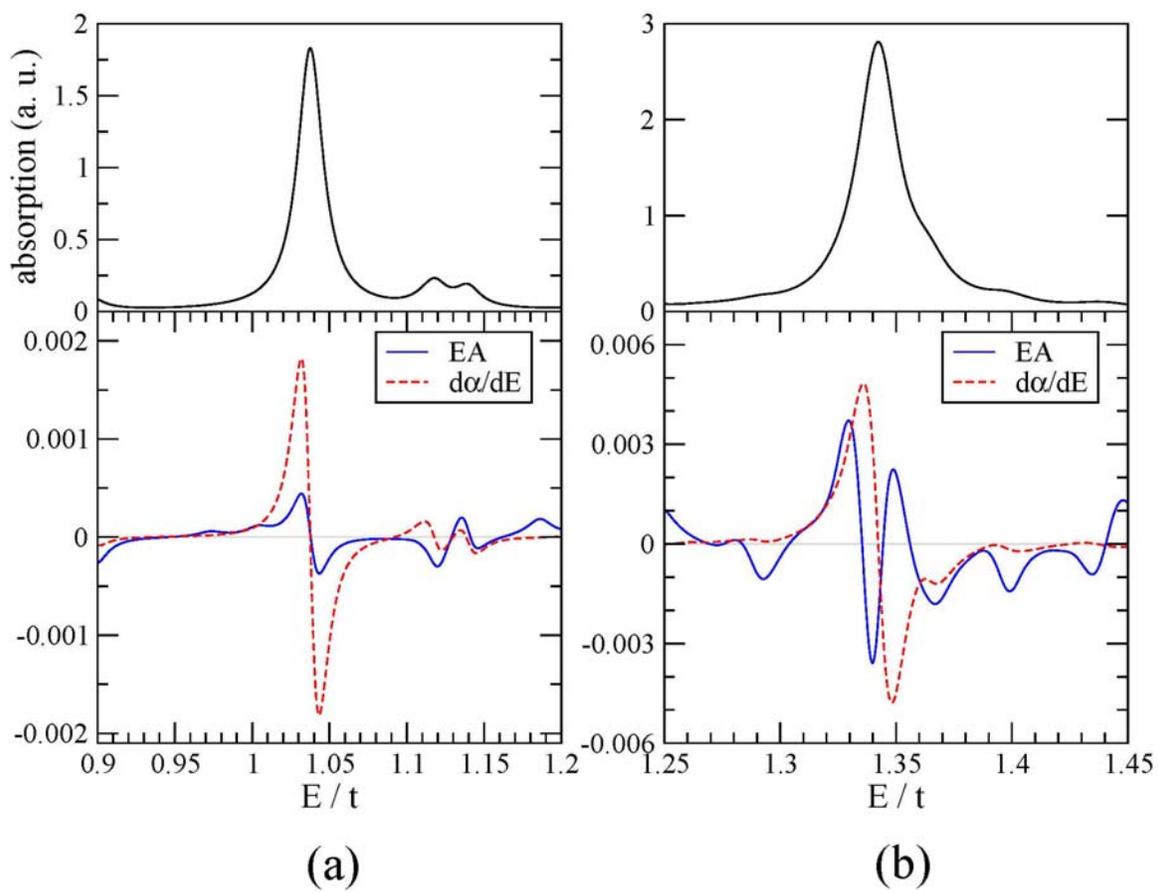

Fig. 6